# Methodology of Parameterization of Molecular Mechanics Force Field From Quantum Chemistry Calculations using Guided Genetic Algorithm: A case study of methanol

Ying Li,[a] Hui Li,[b] Maria K. Y. Chan,[c,d] Subramanian Sankaranarayanan,[c] Benoît Roux[b,d]

[a] Leadership Computing Facility, Argonne National Laboratory, IL 60439, USA
[b] Department of Biochemistry and Molecular Biophysics, University of Chicago, IL 60637, USA
[c] Computational Institute, University of Chicago, IL 60637, USA
[d] Center for Nanoscale Materials, Argonne National Laboratory, IL 60439, USA

**ABSTRACT**

In molecular dynamics (MD) simulation, force field determines the capability of an individual model in capturing physical and chemistry properties. The method for generating proper parameters of the force field form is the key component for computational research in chemistry, biochemistry, and condensed-phase physics. Our study showed that the feasibility to predict experimental condensed phase properties (*i.e.*, density and heat of vaporization) of methanol through problem specific force field from only quantum chemistry information. To acquire the satisfying parameter sets of the force field, the genetic algorithm (GA) is the main optimization method. For electrostatic potential energy ($E_{ESP}$), we optimized both the electrostatic parameters of methanol using the GA method, which leads to low deviations of $E_{ESP}$ between the quantum mechanics (QM) calculations and the GA optimized parameters. We optimized the van der Waals (vdW) parameters both using GA and guided GA methods by calibrating interaction energy ($\Delta E$) of various methanol homo-clusters, such as nonamers, undecamers, or tridecamers. Excellent agreement between the training dataset from QM calculations (*i.e.*, MP2) and GA optimized parameters can be achieved. However, only the guided GA method, which eliminates the overestimation of interaction energy from MP2 calculations in the optimization process, provides proper vdW parameters for MD simulation to get the condensed phase properties (*i.e.*, density and heat of vaporization) of methanol. Throughout the whole optimization process, the experimental value were not involved in the objective functions, but were only used for the purpose of justifying models (*i.e.*, nonamers, undecamers, or tridecamers) and validating methods (*i.e.*, GA or guided GA). Our method shows the possibility of developing descriptive polarizable force field using only QM calculations.

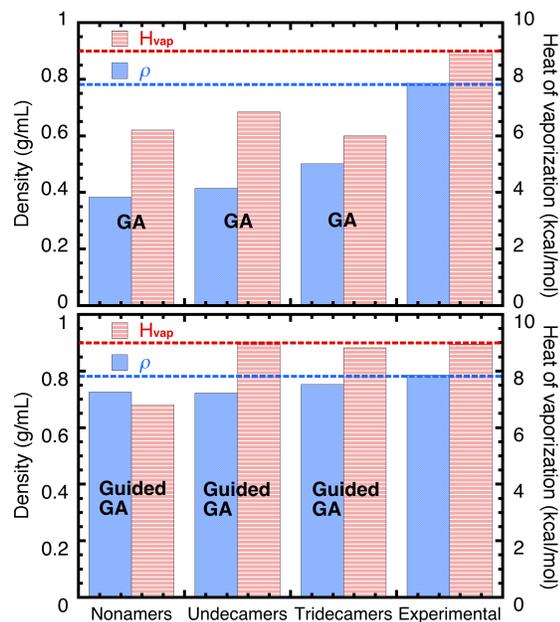

SECTION: Molecular Mechanics, Quantum Chemistry, Force Field and General Theory

INTRODUCTION

Molecular mechanics (MM) modeling have been extensively employed in the research of chemistry, biochemistry, and condensed-phase physics, due to its capability of providing atomistic resolution data within relatively small to moderate computational cost. The accuracies of the MM calculation results relies on several important factors, such as sampling techniques, the underlying function forms and the parameters employed in the potential, with the latter two being the indispensable components of MM force field. Force field determines the capability of a certain model in capturing physical and chemistry properties in molecular dynamics (MD) simulations. The construction of a force field and its parameterization are non-trivial processes that require thoroughly understanding of the underlined physics and chemistry, carefully designing of parameterization protocols and appropriately applying of the optimization techniques. While several force field parameterization strategies have been widely employed and well-documented elsewhere,[1] these approaches rely more or less on both experimental and quantum mechanics (QM) calculated properties. The present study intends to extend the existing scopes of force field developing strategies, to explore the possibility to use only QM calculation approaches for force field parameterization.

In this study, we probed the procedure for parameterization of the polarizable force field in the form of Atomic Multipole Optimized Energetics for Biomolecular Applications (AMOEBA) based on QM calculations. Comparing with non-polarizable force fields, which describes the system dipole moments in an averaged and thus fixed fashion, polarizable models were designed in a way, so that they can effectively describe the capability of chemistry species to change electronic distribution under the influence of its electrostatics environment.[2] The polarizable models represented a significant improvement in the field of theoretical chemistry since last decade and had led to important scientific insights such as in the study of ionic liquids and ligand binding to biological macromolecules.[3] Knowledge on efficient and accurate parameters development of polarizable models will lead to more discoveries in biophysics, biochemistry and material science. There are several polarizable force fields available for simulating various chemical and biomolecular systems, such as the Drude polarizable force field that utilizes charges on springs,[2a, 2b, 4] the CHARMM charge equilibration (CHEQ) model that employs fluctuating charges,[5] the AMBER ff02 using partial charges with inducible dipoles,[6] and the AMOEBA force field that incorporates contributions from monopoles (charges), dipoles and quadrupoles, that are of higher order multipoles.[2c, 7] The AMOEBA polarizable force field has illustrated its capability of providing reasonable predictions of interaction energy and structural properties of small molecules and clusters in gas phase, as well as the physicochemical properties of bulk phase, such as density, self-diffusivity and static dielectric constants.[3e]

In order to improve the ability to describe the physical models, AMOEBA polarizable force field invokes the utilization of a vast number of variables at the atomic site. The increments of the dimension of variables lead to a series of challenges in parameter searching and thus have been a focus of the present study. To address this issue, we implement the genetic algorithm (GA) to approach the global optimization regarding parameterizations. GA is an evolutionary algorithm that mimics the process of natural selection.[8] For traditional parameters optimization, gradient-based algorithms, such as the steepest descent method, conjugate gradient method, *etc.* are the standard approach.[9] However, for the gradient-based method, the optimal result depends strongly on the existing knowledge of the objective function. For example, the convergence of the optimal parameters is not guaranteed when the surface of the objective function is considerably rugged or non-differentiable, or when the initial value deviates significantly from the global minimum. In parameters searching of force field, the value range of parameters exponentially expands as the complexity of the force field functional form increases. To tackle the optimization algorithm issue, statistical search heuristic (such as GA, simulated annealing, *etc.*) is much needed to drive the parameter space towards the optimal region for the objective function. Some successful examples are using the GA to determine the parameters for the reactive force field (ReaxFF) and hybrid bond-order potential (HyBOP) for materials system.[10]

Another issue to address in this study is the selection of reference data. Force fields that have been extensively employed and validated are often based on reference data from both experiments and theoretical calculations. For example, the development of almost all the widely employed water models[2c, 11] are based on achieving a balance between the high level *ab initio* QM calculated dataset and condensed phase properties of bulk water measured by experimentalists. Calibration of force field parameters can be achieved by minimizing the difference of the desired properties between the MM calculations and QM calculations/experimental data.[12] For small molecules and clusters in gas phase, QM calculations are often straightforward and within reasonable computational cost. While the evaluation of the bulk phase properties from QM calculations is often computationally unfeasible, thus referring to experimental data (such as densities, dielectric constants, *etc.*) is necessary. However, for many molecules and especially mixtures system, experimental data are either unavailable or have significant discrepancies between different studies, due to difficulties in achieving precise measurements. Thus it is necessary to explore the possibility to select reference data of small molecules or clusters solely from theoretical calculations, *i.e. ab initio* QM calculations, as the training dataset for force field development to make the predictive calculation in computational modeling.

**METHODS**

The principal goal of this work is to establish guidance for constructing a framework of force field parameters development using primarily QM calculation data. In this initial attempt, we selected liquid methanol as the model system. Methanol represents a molecule of significant importance to organic liquids and polymers that contain bulky alkyl moieties. The hydroxyl group leads the molecule to the polar nature. The formation of the methanol clusters primarily comes from the contribution of hydrogen bond networks.[13] These ubiquitous aspects make methanol as an excellent candidate for designing and examination of force field development.

Ren *et. al.* suggested a general protocol of parameter development for the AMOEBA polarizable force field.[7, 14] In the AMOEBA model, the electrostatic potential explicitly considers the effect of atomic multipoles (charge, dipoles, and quadrupoles) and polarization effect on induced dipole. The van der Waals (vdW) term is modeled as a buffered 14-7 potential proposed by Halgren.[15] Compared to the 12-6 potential, the buffered 14-7 potential has the advantage of optimizing structures with initial crude geometries. For hydrogen atoms, an additional parameter (the so-called reduction factor) is used to scale the position of the hydrogen atom interacting site along the corresponding covalent bond. It is meant to reflect the degree of which hydrogenic electron density displaced toward the heavy atom when covalent bonding takes place.[3e] The full potential energy function of the AMOEBA force field has been described in full detail elsewhere.[2c, 14] The general energy expression of the AMOEBA force field is written in equation (1), where parameterizations of electrostatic potential energy $E_{ESP}$ and vdW interaction energy $E_{vdW}$ terms are studied in this work. Bonded interaction, including bond, angle, dihedral and torsional parameters, were kept consistent with that in the original AMOEBA force field (amoeba09.prm).[7]

$$\begin{aligned} E_{AMOEBA} &= E_{bonded} + E_{no-bonded} \\ &= E_{bonded} + E_{ESP} + E_{vdW} \end{aligned} \quad (1)$$

In Ren *et. al.*'s protocol, the parameters of multipoles and torsional bonded terms are optimized using QM results. However, besides of taking QM calculations as the reference, the protocol adapts data from existing database of experimental values, which are not entirely available for every desired system. We followed their approaches with the context of parameterizing standard AMOEBA force field for atomic multipoles, polarizabilities and vdW interaction by GA methods using solely results from QM calculations, which are available for the methanol system within given computational resources.

All the QM calculations were performed with Gaussian 09[16] package of electronic structure programs. The second order Møller–Plesset perturbation theory[17] (MP2) was employed, with the basis set superposition error (BSSE) correction.[18] This level of theory and basis set to predict energetics and structural properties has been widely verified by previous studies.[19] The electronic density calculation of methanol monomer is initially used for calibrating the magnitude of the multipole moments at the atomic positions. Combining with the Distributed Multipole Analysis (GDMA 2.2) tool developed by Anthony Stone,[20] the Tinker[2c, 21] package is used to optimize the permanent atomic multipoles by fitting the electrostatic potential measured on the Connolly surface of the methanol monomer. Table S1 in the Supporting Information[22] is showing the electrostatic potential results for methanol monomer using different basis sets.

In the parameterization process, the training dataset for the optimization should contain a good sampling of the possible structural configurations and their respective energies. For instance, to predict the polarizable effect between molecules, methanol dimers with a continuous range of electrostatic potential energies ($E_{ESP}$) should be included. In AMOEBA, to compute clusters electrostatic potential energy ($E_{ESP}$) including the polarizable effect between methanol molecules, 44 independent parameters are needed. Those parameters are monopole ($q$), dipole ($\mu_x$, $\mu_y$, $\mu_z$), quadrupole-a traceless and systemic matrix- ($Q_{xx}$, $Q_{yx}$, $Q_{yy}$, $Q_{zx}$, $Q_{zy}$), atomic polarizability ($\alpha$) and Thole's description of damping factor ($a$) for four types of atoms (O, H(-O), C and H(-C)) in methanol (The detailed description of the parameter form can be found in somewhere else[2c, 14]). We optimized all 44 independent parameters listed above, using an extensive training dataset taking electronic density calculations for 4943 methanol dimers. The methanol dimer configurations were optimized via Guassian09 program at the MP2 level using 6-311G(d, p) basis set. The electrostatic potential energy ($E_{ESP}$) of 4943 dimers was computed using the same level of theory and basis set. The detailed description of optimization for electrostatic parameters is in the Supporting Information Section S2.[22]

To ensure an adequate representation of the various possible coordination environments and cluster sizes as well as the energy landscape, we sampled three training datasets containing different methanol homoclusters, which consisting of 9, 11, and 13 molecules referred as nonamer, undecamer and tridecamer in the descriptions below, respectively, to calibrate the vdW interaction energy ($E_{vdW}$). For methanol, there are vdW potential depth ($\varepsilon$) and minimum energy distance ($R^*$) for four types of atoms, and additional reduction factor ($\lambda$) for two types of hydrogen atoms, which constitutes 10 independent parameters to optimize. We optimized these 10 independent parameters using interaction energy ($\Delta E$) results of the sampled 502 nonamers, 157 undecamers and 94 tridecamers at the MP2/6-31G (d, p), respectively. For instance, the nonamer was sampled through randomly placing 9 C atoms in a 10 Å × 10 Å × 10 Å computational supercell, where one C atom placed in the center and surrounded by other 8 C atoms with relative distance at least of 3.25 Å, the minimum distance of carbon bond networks in methanol,[23] to each other. Then, methanol with random orientations will be assigned to the position where the C atoms sit. The methanol nonamer configuration is optimized in Gaussian09 using the eigenvalue-following algorithm[24] with constraints of fixed C atoms positions. The relaxed nonamer configurations and their interaction energies ($\Delta E$) between the center methanol and surrounding methanols are then employed in the training dataset for optimizing the vdW parameters. This procedure was applied to undecamers and tirdecamers as well for getting the relaxed configuration and the interaction energy, respectively. Figure 1 (a), (b) and (c) are showing example configurations of nonamer, undecamer and tridecamer.

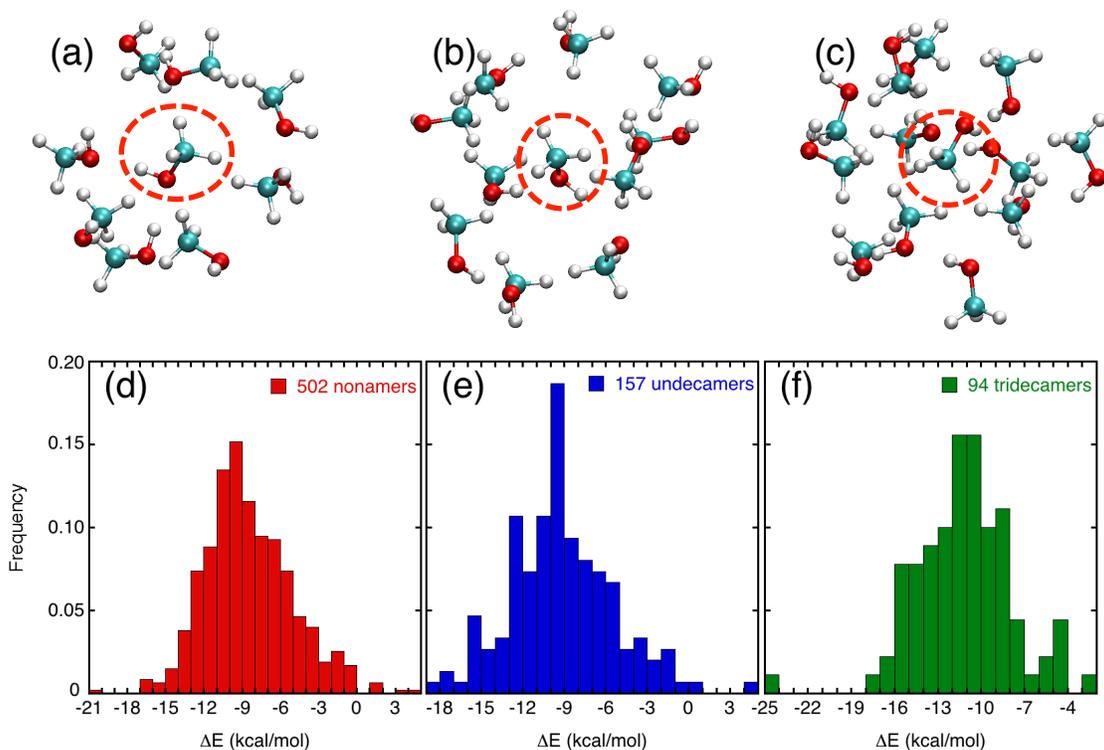

Figure 1. Demonstration of the configuration of a methanol (a) nanomer, (b) undecamer, (c) tridecamer, where H atom in white, C atom in cyan and O atom in read. The center molecule is circled, where the rest are surrounding molecules. The interaction energy of the nonamer is the total energy of the nonamer subtracting the total energy of the center molecule and the surrounding molecules. Distribution of interaction energies of (d) 502 methanol nonamers, (e) 157 undecamers, (f) 94 tridecamers.

In the AMOEBA force field, except for the electrostatic potential energy ($E_{ESP}$), vdW interaction energy ($E_{vdw}$) is the only term consisting of interaction energy between molecules. The vdW interaction energy ($E_{vdw}$) can be calibrated from the interaction energy ($\Delta E$) by separating apart molecules, with the assumption that the deformation energy of each molecule upon binding is reasonably small.[7] Through sampling configurations widely and calculating the coordinated interaction energies, we consider the vdW interaction by taking the effect of electrostatic interaction inclusively. The interaction energy ($\Delta E$) is calculated as the total energy of the cluster subtracts the total energy of the center methanol molecule and that of the surrounding methanol clusters, as shown in equation (2). The BSSE correction is applied for getting the $\Delta E$ of every configuration. The MP2/6-31G (d, p) computed interaction energies of the 502 nonamer, 157 undecamer and 94 tridecamer configurations exhibit a Gaussian distribution as shown in figure 1 (d), (e) and (f), which signify the sampling of those configurations are adequate and in the equilibrated states.

$$\Delta E = E_{cluster} - E_{center} - E_{sourrounding} \qquad (2)$$

Using the training dataset described above, we optimized the non-bonded parameters in AMOEBA force field by employing genetic algorithm; the procedure is outlined in Figure 2.

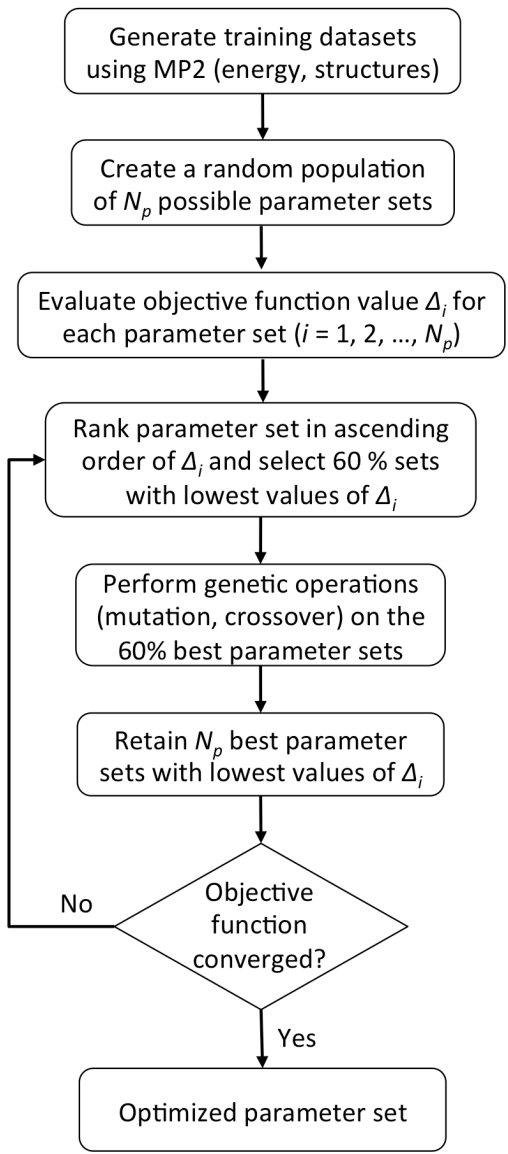

Figure 2. Flowchart describing the sequence of steps employed in this work for optimization of vdW parameters in AMOEBA force field.

Taking the optimization of vdW parameters as an illustration, we begin the optimization process by generating a population of $N_p$ = 120 parameter sets randomly, such that their values lie within physically allowable limits (parameter search ranges are listed in Supporting Information Table S2[22]). Each set of parameters in this population is called a member. For each member $i$, we compute the interaction energies for all structures in the training dataset using MD simulation package Tinker[2c, 21] and evaluate the objective function $\Delta_i$ given by equation (3),

$$\Delta_i = \sum_j \left( \Delta E_j^{MP2} - \Delta E_j^{AMOEBA_i} \right)^2 \qquad (3)$$

where $\Delta E_j^{MP2}$ and $\Delta E_j^{AMOEBA_i}$ are the MP2 calculated and the AMOEBA calcuated interaction energies for the structure $j$ in the training dataset, respectively. Since the electrostatic potential energy ($E_j^{ESP}$) is computed for the structure $j$ to represent the atomic electrostatic potential precisely, the only uncertain term in the interaction energy is the vdW interaction energy ($E_j^{vdW}$), represented by a member $i$, a set of 10 independent parameters. Here, members are then ranked in an ascending order of $\Delta_i$. After the ranking, non-linear roulette wheel selection[25] will be performed to select the top 60% members, *i.e.*, the ones with lowest values of $\Delta_i$, which are then subjected to genetic operations: crossover with crossover-rate 3% and mutation. These mutations introduce sufficient diversity into the population, and the non-linear selection scheme helps to avoid premature convergence of the GA run. After the genetic operations, both the old and the new members are ranked by their $\Delta$s. The best $N_p$ parameter sets (members) were then chosen to constitute the next generation. Such an optimization routine ensures that only satisfactory parameter sets survive after each generation; upon repeating this workflow for sufficient generations and sampling viable regions in the parameters space, we perform three separate GA runs starting with different random populations. From each of the converged GA run, we choose the final parameter set corresponding the lowest Δ.

To overcome one of the serious shortcomings of MP2 theory, a noticeable overestimation of the interaction energy,[26] which plays a major role in stability and formation of condensed phase molecular structure, we improved the form of objective function Δ in the GA program. We introduced a penalty parameter $\delta$ in the GA optimization procedure as the unexplained discrepancy between the AMOEBA results and MP2 calculations, as shown in equation (4), which obeys the design concept of guided genetic algorithm (guided GA).[27] This penalty parameter $\delta$ never participates as the form of AMOEBA force field in MD simulations, adds up to 11 independent parameters as a subset of $i^{th}$ member in the GA optimization process for vdW parameters.

$$\Delta_i = \sum_j \left( (\Delta E_j^{MP2} - \delta) - \Delta E_j^{AMOEBA_i} \right)^2 = \sum_j \left( \Delta E_j^{MP2} - (\Delta E_j^{AMOEBA_i} + \delta) \right)^2 \qquad (4)$$

**RESULTS**

The vdW parameters are critical in the force field for getting the correct condensed phase structures, such as right intermolecular distance, through molecular dynamic simulations. In this section, we present the optimization results of the vdW parameters from both GA and guided GA program by calibrating interaction energy between molecules. We follow with showing binding curves for methanol dimers using both GA and guided GA optimized parameters. Finally, we show the MD simulation results (*i.e.*, density and heat of vaporization) of a condensed phase methanol system from both GA and guided GA optimized parameters using AMOEBA force field.

**3.1. VDW parameters of nonamers, undecamers, and tridecamers from GA and Guided GA**

The GA and guided GA programs were applied to optimize the vdW parameters using the interaction energies ($\Delta E$) calculated at MP2/6-31G (d, p) with BSSE correction for 502, 157 and 94 methanol nonamers, undecamers and tridecamers, respectively. The excellent correlations (R = 0.970, 0.976, and 0.968) of the $\Delta E$ between MP2 calculations and AMOEBA with GA optimized parameters for nonamers, undecamers, and tridecamers, respectively, are shown in figure 3 (a), (b) and (c). Meanwhile with apparent overestimation of the MP2 calculation, figure 3 (d), (e) and (f) are also showing excellent correlations (R = 0.956, 0.965 and 0.946) of the $\Delta E$ between MP2 calculations and AMOEBA with guided GA optimized parameters for nonamers, undecamers, and tridecamers, respectively. There are in total six sets of vdW parameters from nonamers, undecamers, and tridecamers all using GA and guided GA optimization method, respectively.

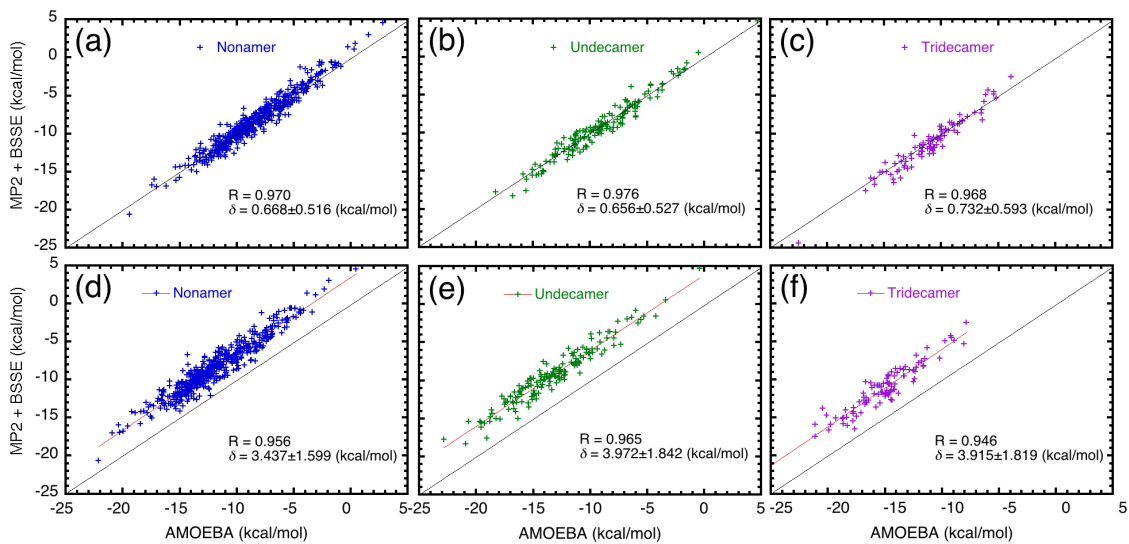

Figure 3. Comparison of the interaction energy ($\Delta E$) for (a) 502 methanol nonamers, (b) 157 methanol undecamers, (c) 94 methanol tridecamers computed from MP2/6-31G(d, p) + BSSE calculations and GA optimized AMOEBA model. Comparison of the interaction energy ($\Delta E$) for (d) 502 methanol nonamers, (e) 157 methanol undecamers, (f) 94 methanol tridecamers computed from MP2/6-31G(d, p) + BSSE calculations and guided GA optimized AMOEBA model.

### 3.2. VDW parameters for Dimers

To validate the transferability of the GA and guided GA optimized AMOEBA parameters from nonamers, undecamers, and tridecamers for gas-phased methanol molecules, we use the six sets of parameters to calibrate the binding energy curves of methanol dimers. We fully optimized 141 methanol dimer configurations at MP2/6-31G (d, p) with BSSE correction with the constraint of pre-defined distance between carbon atoms in the two methanol molecules. Figure 4 is showing the binding energy of methanol dimers as a function of the distance between two carbon atoms.

We obtained the binding energy curves for the 141 configurations of methanol dimer using the three sets of vdW parameters from nonamers, undecamers, and tridecamers using GA optimization method, respectively. Figure 5 (a), (b) and (c) are showing the good agreement of the binding energy between the MP2 calculation and the GA optimized parameters from nonamers, undecamers, and tridecamers, respectively. We also obtained the binding energy curves for those dimers using the other three sets of vdW parameters using guided GA optimization method from nonamers, undecamers, and tridecamers. Figure 5 (d), (e) and (f) are showing the binding energy curves of methanol dimers from the MP2 calculations and from the guided GA optimized vdW parameters by nonamers, undecamers, and tridecamers, respectively. From the lower row of figure 5, we can see discrepancies of binding energy between the gas-phased MP2 calculations and the guided GA optimized parameters from clusters, where the overestimation of MP2 calculations is conspicuous near the lowest energy curve.

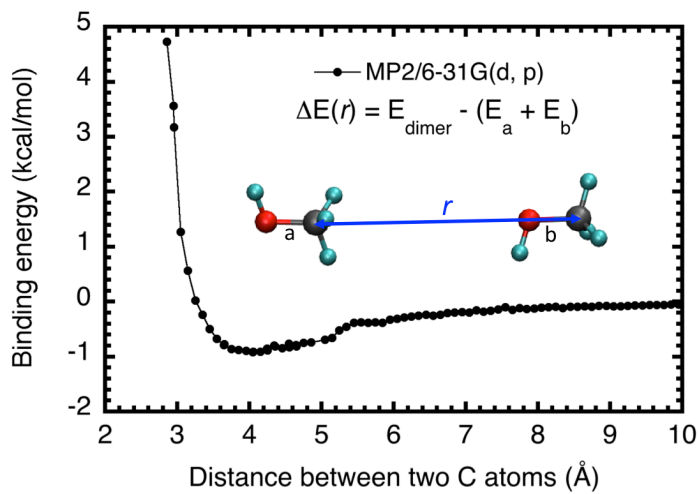

Figure 4. MP2/6-31G (d, p) + BSSE calculated binding energy of methanol dimer as a function of the distance between the carbon atom of each methanol.

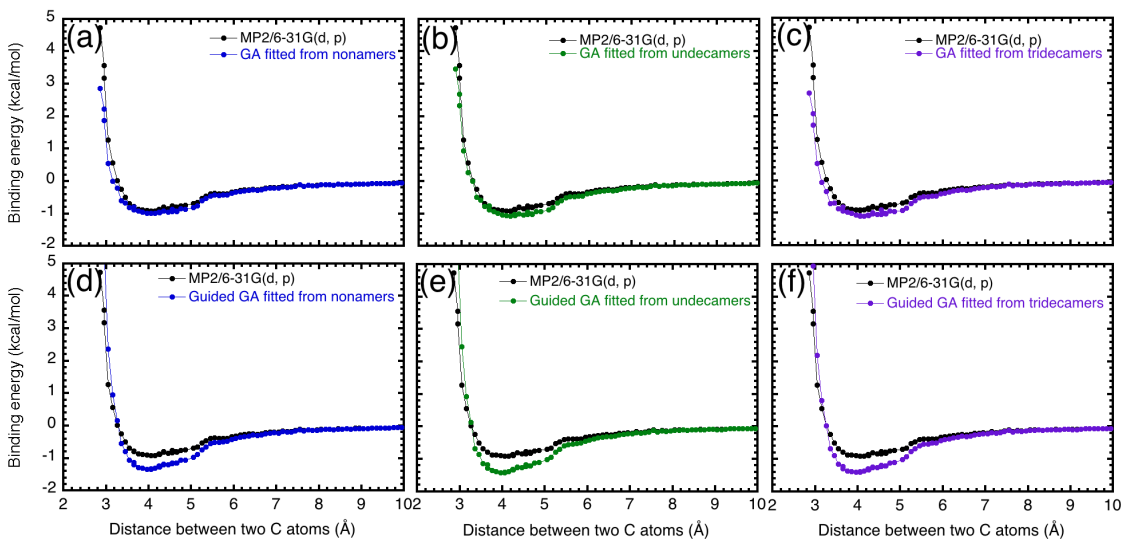

Figure 5. Comparison of binding energy curves of methanol dimer from MP2 calculation and GA optimized AMOEBA model using (a) nonamers, (c) undecamers, (f) tridecamers. Comparison of binding energy curves of methanol dimer from MP2 calculation and guided GA optimized AMOEBA model using (b) nonamers, (d) undecamers, (f) tridecamers.

The shifting down of the binding energy curves from the guided GA optimized parameters implies the interaction energy of the same level of QM theory (*i.e.*, MP2/6-31G(d, p) + BSSE) calculations from gas-phased molecules leads to the overestimation of real potential energy surface (PES). This discrepancy of interaction energy from MP2 calculations is commonly seen as the incompleteness of the basis sets for the particular level of MP2 theory and the insufficiency sampling of configurations in the parameterization of force field.[19, 26]

### 3.3 VDW parameters for Condensed Phase Methanol

To verify the ability of our parameters of predicting the condensed phase properties, sets of MD simulations using the developed parameters of AMOEBA force field were performed. Here, we applied the six sets of optimized parameters to perform MD simulations to the condensed phase system consisting of 344 methanol molecules. For getting the density ($\rho$) and heat of vaporization ($H_{vap}$), we used Tinker package (omm_dynamics program -- GPU version of Tinker dynamics program[28] ) to run MD simulation. NPT ensemble simulations were performed at $T$ = 298.15 K and $P$ = 40 atm with an integration time step of 1 fs. With three different random seeds for initiating velocity on the system, the averaged value of density and heat of vaporization were obtained. Table 1 shows the result of density and heat of vaporization from Tinker MD simulations with the GA optimized AMOEBA force field parameters. Among all the results, the simulated density from tridecamers is the best, while the winner of the heat of vaporization is from undecamers. However, from Table 1, it is clear that the value of density and heat of vaporization are nowhere close to the experimental value.

Table 1. The density and heat of vaporization calculated from Tinker MD simulation with the GA optimized AMOEBA force field from 502 nonamers, 157 undecamers, and 94 tridecamers, respectively.

| Number of methanol in cluster | Number of configurations | Properties | |
|---|---|---|---|
| | | $\rho$ (g/mL) | $H_{vap}$ (kcal/mol) |
| 9 | 502 | 0.383 | 6.202 |
| 11 | 157 | 0.415 | 6.839 |
| 13 | 94 | 0.501 | 5.995 |
| Experimental[29] | | 0.786 | 8.950 |

Table 2 shows the results of density and heat of vaporization from Tinker MD simulations with the guided GA optimized AMOEBA force field parameters. The simulated density and heat of vaporization from tridecamers are much closer to the experimental values. We can see that using the guided GA, which introduces a penalty parameter $\delta$, eliminates the overestimation of interaction energy from MP2 calculation. The guided GA optimized parameters describe better condensed phase properties (*i.e.*, density and heat of vaporization) than the GA optimized parameters in MD simulation. Among all the results, the simulated density and heat of vaporizations from more number of methanol molecule clusters are better. MP2 calculations of interaction energy of tridecamers and undecamers are more descriptive than nonamers for optimizing vdW parameters, due to more detailed information can be calibrated as more molecules in the solvation shell of methanol involved.

Table 2. The density and heat of vaporization calculated from Tinker MD simulation with the guided GA optimized AMOEBA force field from 502 nonamers, 157 undecamers, and 94 tridecamers, respectively.

| Number of methanol in cluster | Number of configurations | Properties | |
|---|---|---|---|
| | | $\rho$ (g/mL) | $H_{vap}$ (kcal/mol) |
| 9 | 502 | 0.726 | 6.807 |
| 11 | 157 | 0.722 | 9.014 |
| 13 | 94 | 0.753 | 8.825 |
| Experimental[29] | | 0.786 | 8.950 |

## CONCLUSIONS

Our study showed that the feasibility to predict experimental condensed phase properties (*i.e.*, density and heat of vaporization) of methanol through problem specific (here, we use AMOEBA) force field from only quantum chemistry information. To acquire the satisfying parameter sets, the genetic algorithm (GA) is the main optimization method. For electrostatic potential energy ($E_{ESP}$), we optimized the both the multipoles and polarizability, Thole damping factors of methanol using the GA method, which leads to lower deviations of $E_{ESP}$ between QM calculations and the GA optimized parameters than that from amoeba09.prm. We optimized the vdW parameters both using GA and guided GA methods by calibrating interaction energy ($\Delta E$) of various methanol homo-clusters, such as nonamers, undecamers, or tridecamers. Excellent agreement between the training dataset from MP2 calculations and GA optimized parameters can be achieved. However, only the guided GA method, which eliminates the overestimation of interaction energy from MP2 calculations in the optimization process, provides proper vdW parameters for MD simulation to get condensed phase properties (*i.e.*, density and heat of vaporization) of methanol. Throughout the whole optimization process, the experimental value were not involved in the objective functions, but were only used for the purpose of justifying models (*i.e.*, nonamers, undecamers, or tridecamers) and validating methods (GA or guided GA). We conclude that the main difficulty of parameterizing the force field for liquid phase methanol from solely QM calculations could be coming from the reliability of the training dataset (*i.e.*, MP2 calculations), as it is well known that different basis sets of MP2 could lead to different accuracy of QM calculations, the conspicuous overestimation of interaction energy from MP2 calculations. Some of the other higher level of QM calculations (*e.g.*, MP3/4, CCSD(T), *etc.*) can only deal with up to certain amount of atoms/electrons.[30] It is ambiguous to study the property of condensed phase properties from a few molecules. In fact, we tried to use the interaction energy ($\Delta E$) calculated for methanol tridecamers from another type of QM calculation, Symmetry Adapted Perturbation Theory (SAPT),[31] as the training dataset. We got very reasonable correlations of the interaction energy between the SAPT calculation and the GA/guided GA optimized vdW parameters, but the yielded condensed phase properties from MD simulations are non-ideal as shown in the Supporting Information Section S3.[22] However, our method was able to utilize the QM calculations (*i.e.*, MP2 calculations) from the limited calculation resource to tune the force field parameters for reasonable agreement between the MD simulated values and experimental data. The method is straightforward to implement and has the potential to be extended to any type of small organic molecule systems, as well as other descriptive polarizable force field.

**ACKNOWLEGDMENTS**
This work was supported by LDRD at Argonne National Laboratory Grant No. XXX. Simulations were performed on Blues computers at the Laboratory Computing Resource Center and Cooley computers at the Leadership Computing Facility at the Argonne National Laboratory, and on the Midway computers of the Research Computing Center at the University of Chicago. We appreciate the Margaret Butler Postdoctoral Fellowship at Argonne Leadership Computing Facility for supporting the work. We thanks to Dr. Frank C. Pickard IV and Dr. Bernard R. Brooks from the Computational Biophysics Section of the Laboratory of Computational Biology at the National Institutes of Health for the very useful discussion on calculation on SAPT method.

Supporting Information for "Methodology of Parameterization of Molecular Mechanics Force Field From Quantum Chemistry Calculations using Guided Genetic Algorithm: A case study of methanol"

Ying Li,[a] Hui Li,[b] Maria K. Y. Chan,[c,d] Subramanian Sankaranarayanan,[c] Benoît Roux[b,d]

[a] Leadership Computing Facility, Argonne National Laboratory, IL 60439, USA
[b] Department of Biochemistry and Molecular Biophysics, University of Chicago, IL 60637, USA
[c] Computational Institute, University of Chicago, IL 60637, USA
[d] Center for Nanoscale Materials, Argonne National Laboratory, IL 60439, USA

**S1. Initial Mutiploes**

To describe the electrostatic properties of a methanol monomer, we first determine the multipole, including the monopole $q$, dipole $\vec{\mu}$, and quadrupole $\vec{\vec{Q}}$ components at atomic sites according to the electron distribution. Electrostatic potential energy ($E_{ESP}$) of an optimized methanol monomer is determined using the MP2 theory with various basis sets including Pople-style[32] and correlation consistent[33] basis sets, as shown in Table S1. We then perform the DMA[20a, 34] analysis as implemented in the GDMA (Gaussian Distributed Multipole Analysis) program[35] for the electronic density results from different MP2 basis sets. Using Tinker[2c, 21] package (the Poledit and Potential program), we identify the multipole parameters of each atomic site.

Table S1. The electrostatic potential energy ($E_{ESP}$) for methanol monomer from MP2 calculations using different basis sets and from the corresponding fitted multipole parameters, with the corresponding root mean square deviation and relative error.

| Basis | ESP from MP2 (kcal/mol) | ESP from fitted multipole (kcal/mol) | RMSD (kcal/mol) | Relative Error (%) |
|---|---|---|---|---|
| 6-31G(d, p) | 4.878 | 4.865 | 0.165 | 0.267 |
| 6-31+G* | 5.675 | 5.649 | 0.172 | 0.458 |
| 6-31G* | 5.059 | 5.046 | 0.170 | 0.257 |
| 6-311G | 5.947 | 5.935 | 0.172 | 0.202 |
| 6-311G(d, p) | 4.815 | 4.800 | 0.148 | 0.312 |
| 6-311G(2df, 2pd) | 4.462 | 4.447 | 0.154 | 0.336 |
| 6-311G* | 5.187 | 5.172 | 0.156 | 0.289 |
| 6-311+G* | 5.648 | 5.626 | 0.167 | 0.390 |
| 6-311++G** | 5.307 | 5.282 | 0.165 | 0.471 |
| 6-311+G** | 5.309 | 5.285 | 0.166 | 0.452 |
| Aug-CC-pvDz | 4.758 | 4.729 | 0.177 | 0.609 |
| Aug-CC-pvTz | 4.748 | 4.723 | 0.165 | 0.527 |
| Aug-CC-pvQz | 4.745 | 4.720 | 0.165 | 0.527 |

Note that the multipoles parameter based on the MP2 calculations using a larger basis set does not necessary yields more accurate $E_{ESP}$. For instance, the result from 6-311G(d, p) basis set is closer to the Aug-CC-pvD/T/Qz basis sets than the result from 6-311G(2df, 2pd). The root mean square derivation (RMSD) of $E_{ESP}$ is calculated as shown in equation (S1):

$$RMSD_i^{ESP} = \sqrt{\frac{1}{n^{gird}} \sum_{k=1}^{n^{grid}} \left(\phi_k^{MP2} - \phi_k^{AMOEBA_i}\right)^2} \qquad (S1)$$

where $n^{grid}$ is the number of grid points on the monomer's Connolly surface at where $E_{ESP}$ are calculated, and $\phi_k^{MP2}$ and $\phi_k^{AMOEBA}$ are the $E_{ESP}$ calculated at the $k^{th}$ point from MP2 and AMOEBA, respectively. The initial atomic multipoles parameters variance for methanol monomer from various basis sets is considerably small.

## S2. Optimal Multipoles, Polarizability And Damping Factor

In order to predict the polarizable effect between molecules, we used 4943 configurations of methanol dimer and their $E_{ESP}$ to parameterize the optimal electrostatic parameters (i.e., multipoles ($q, \vec{\mu}, \ddot{Q}$), atomic polarizability ($\alpha$) and Thole's description of damping factor ($a$)). The methanol dimer configurations were sampled through placing two methanol molecules, where one methanol was sampled over shell radius (1- 4 Å) on another methanol's Connolly surface. In total, 4943 methanol dimer configrations were sampled. These 4943 configurations of methanol dimers are relaxed through MP2 geometry optimization with constraint of fixing position of carbon atoms. We adopted ±50 % more of the magnitude of atomic multipoles fitted from methanol monomer at MP2/6-311G (d, p) from Section S1 as the electrostatic parameters searching range seen in Table S2. Then for $i^{th}$ member of set parameters, we use the GA program by minimizing the objective function $\Delta_i$ as the averaged root mean square deviation (ARMSD) between the MP2/6-311G (d, p) result and the parameterized AMOEBA calculation, as shown in equation (S2):

$$\Delta_i = ARMSD_{dimer}^{ESP} = \frac{1}{N}\sum_{j=1}^{N} RMSD_j^{ESP_i} = \frac{1}{N}\sum_{j=1}^{N}\left(\sqrt{\frac{1}{n^{gird}}\sum_{k=1}^{n^{grid}}\left(\phi_k^{MP2} - \phi_k^{AMOEBA_i}\right)^2}\right)_j^{ESP} \qquad (S2)$$

where $N$ = 4943, $n^{grid}$ is the number of grid points on the dimer's Connolly surface at where $E_{ESP}$ are calculated, and $\phi_k^{MP2}$ and $\phi_k^{AMOEBA}$ are the $E_{ESP}$ calculated at the $k^{th}$ point from MP2 and AMOEBA, respectively.

Table S2. Range of the 44 independent electrostatic parameters and the 11 independent vdW parameter over which the GA optimization were performed.

| Parameters | | Atom types | | | |
|---|---|---|---|---|---|
| | | O | H (-O) | C | H (-C) |
| $E_{ESP}$ | Monopole ($q$) | $-1.0 \sim 0$ | $0 \sim 0.4$ | $0 \sim 0.3$ | $0 \sim 0.06$ |
| | Dipole ($\mu_x, \mu_y, \mu_z$) | $(0 \sim 0.5, 0.0, 0 \sim 0.4)$ | $(-0.14 \sim 0, 0.0, -0.4 \sim 0)$ | $(-0.4 \sim 0, 0.0, 0 \sim 0.9)$ | $(-0.1 \sim 0, 0.0, -0.1 \sim 0)$ |
| | Quadrupole $\begin{bmatrix} Q_{xx} & * & * \\ Q_{yx} & Q_{yy} & * \\ Q_{zx} & Q_{zy} & * \end{bmatrix}$ | $\begin{bmatrix} 0 \sim 0.6 & * & * \\ 0.0 & -0.8 \sim 0 & * \\ 0 \sim 0.06 & 0.0 & * \end{bmatrix}$ | $\begin{bmatrix} 0 \sim 0.28 & * & * \\ 0.0 & 0 \sim 0.06 & * \\ -0.2 \sim 0 & 0.0 & * \end{bmatrix}$ | $\begin{bmatrix} 0 \sim 0.06 & * & * \\ 0.0 & -0.11 \sim 0 & * \\ -0.6 \sim 0 & 0.0 & * \end{bmatrix}$ | $\begin{bmatrix} 0 \sim 0.15 & * & * \\ 0.0 & -0.17 \sim 0 & * \\ -0.08 \sim 0 & 0.0 & * \end{bmatrix}$ |
| | Polarizability ($\alpha$) | $0.5 \sim 1.0$ | $0.2 \sim 0.7$ | $1.0 \sim 1.6$ | $0.2 \sim 0.7$ |
| | Thole's factor ($a$) | $0.3 \sim 0.5$ | $0.3 \sim 0.5$ | $0.3 \sim 0.5$ | $0.3 \sim 0.5$ |
| $E_{vdW}$ | Minimum energy depth ($\varepsilon$) | $3.5 \sim 4.0$ | $2.5 \sim 3.0$ | $2.2 \sim 2.8$ | $3.5 \sim 4.0$ |
| | Minimum energy distance ($R^*$) | $0.12 \sim 0.15$ | $0.001 \sim 0.0015$ | $0.4 \sim 0.6$ | $0.001 \sim 0.009$ |
| | H reduction factor ($\lambda$) | N/A | $0.9 \sim 0.95$ | N/A | $0.9 \sim 0.95$ |
| | Penalty ($\delta$) | $0.05 \sim 8.0$ | | | |

By parameterizing the 9 atomic multipoles ($q, \vec{\mu}, \ddot{Q}$), 1 polarizability ($\alpha$) and 1 damping factor ($a$) through GA program for different types of atoms in methanol (*i.e.*, O, H(-O), C and H(-C)), instead of adapting the values from Thole,[36] we were able to obtain the ESP from AMOEBA as close as the $E_{ESP}$ from MP2 calculations, with even smaller ARMSD value than that given by the amoeba09.prm force field, as shown in Table S3. In order to validate the 44 independent parameters optimized from the GA program, we also calculated the $E_{ESP}$ of 502 configurations of methanol nonamer (a cluster of 9 methanol molecules). Comparing with the ARMSD of $E_{ESP}$ between MP2 calculations and amoeba09.prm force field results, the lower ARMSD value between MP2 calculations and the GA optimized parameters results is shown in Table S3.

Table S3. Averaged root mean square deviation (ARMSD) of electrostatic potential energy ($E_{ESP}$) between MP2 calculations and amoeba09.prm vs. the ARMSSD of $E_{ESP}$ between MP2 calculations and the GA optimized parameters for 4943 methanol dimers and 502 methanol nonamers. All MP2 calculations were performed using 6-311G(d, p) basis sets.

| ARMSD of $E_{ESP}$ (kcal/mol) | amoeba09.prm | GA optimized |
|---|---|---|
| 4943 dimers | 0.723 | 0.305 |
| 502 nonamers | 0.718 | 0.495 |

## S3. vdW parameters From SAPT Method

To justify our GA and guide GA methods used for optimization of the vdW parameters are unique to the MP2 calculated dataset. We also investigated the calculation on the interaction energy ($\Delta E$) using the Symmetry Adapted Perturbation Theory (SAPT)[31] calculations for methanol tridecamers, since it looks like to us that the vdW interaction would involve many body interactions, which means more molecules interaction will give better/more sophisticated potential surface. Figure S1 (a) and (b) are showing the excellent correlations (R = 0.980 and 0.976) between SAPT calculations and the GA and guided GA optimized vdW parameters, respectively. Even our GA and guided GA methods perform well in the optimization process for getting the vdW parameters, the yielded condensed phase properties are not ideal shown in Table S4.

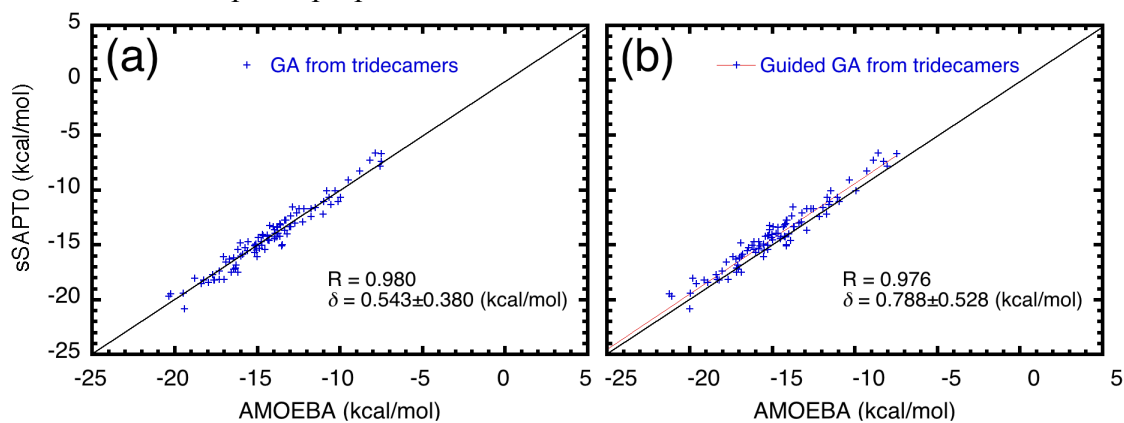

Figure S1. Comparison of the interaction energy ($\Delta E$) for 94 methanol tridecamers computed from sSAPT0/jun-cc-pvDz calculation and (a) GA optimized , (b) guided GA AMOEBA model.

Table S4. The density and heat of vaporization calculated from Tinker MD simulation with the GA and guided GA optimized AMOEA force field from sSAPT0 calculations of 94 tridecamers, respectively.

| Prosperities | | $\rho$ (g/mL) | $H_{vap}$ (kcal/mol) |
|---|---|---|---|
| Optimization method | GA | 0.678 | 5.001 |
| | Guided GA | 0.719 | 5.103 |
| Experimental[29] | | 0.786 | 8.950 |

We conclude because the SAPT calculated interaction energy ($\Delta E$) couldn't be directly expressed as an ubiquitous overestimation/underestimation of real potential energy surface (PSE), which is believed can be calculated using the "gold standard" CCSD(T) method,[30c] this leads GA methods yields bad performance granted. And the design of the guided GA should not be just using one penalty parameter $\delta$, which may be complicated enough and contrary to the intention of our study: a case study on methanol of methodology for polarizable force field development. However, by observing from the binding energy curves calculation for water dimer system as shown in Figure S2, the absolute discrepancy between the SAPT calculation and the real PSE are obviously less than the MP2/6-31G calculation (also shown in less $\delta$ value in guided GA optimization comparing with MP2 calculations). The less absolute discrepancy hints some advantageous of using SAPT method for force field optimization in some other ways.

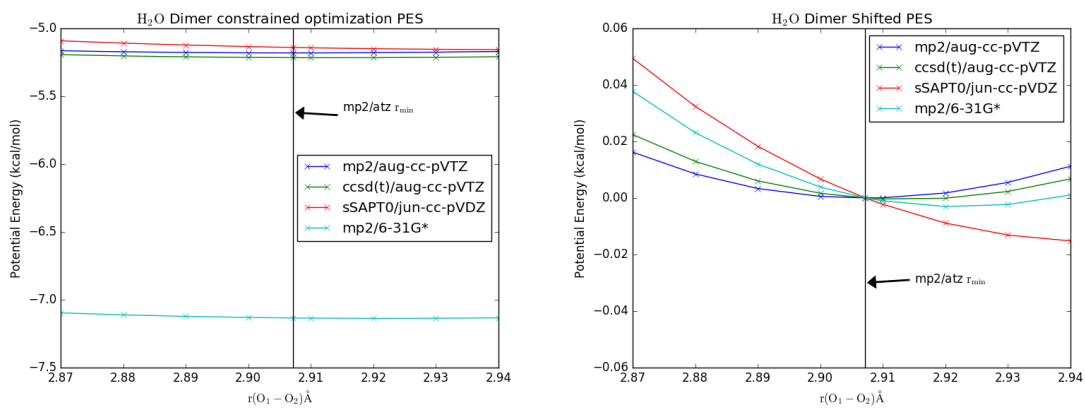

Figure S2. (a) Binding energy curves calculation from different methods for water dimer system, (b) shifted binding energy curves for detailed look of the real PSE, where the zero energy reference point is where the minimum energy calculated from MP2/6-31G.